\documentclass[pra,aps,twocolumn,10pt,showpacs,groupedaddress,superscriptaddress,floatfix,notitlepage]{revtex4-2}
\usepackage{amsmath,amsfonts,amssymb,graphics,graphicx,epsfig,color,times, braket}
\usepackage[utf8x]{inputenc}
\usepackage{color}
\usepackage{bbm, dsfont}
\usepackage{subfigure}
\usepackage{hyperref}
\usepackage{mathrsfs}
\usepackage{verbatim}
\begin{document}
\title{Atomic excitation trapping in dissimilar chirally-coupled atomic arrays}

\author{I Gusti Ngurah Yudi Handayana}
\email{ngurahyudi@unram.ac.id}
\affiliation{Molecular Science and Technology Program, Taiwan International Graduate Program, Academia Sinica, Taiwan}
\affiliation{Department of Physics, National Central University, Taoyuan City 320317, Taiwan}
\affiliation{Institute of Atomic and Molecular Sciences, Academia Sinica, Taipei 10617, Taiwan}

\author{Chun-Chi Wu}
\affiliation{Institute of Atomic and Molecular Sciences, Academia Sinica, Taipei 10617, Taiwan}
\author{Sumit Goswami}
\affiliation{Institute of Atomic and Molecular Sciences, Academia Sinica, Taipei 10617, Taiwan}
\author{Ying-Cheng Chen}
\affiliation{Institute of Atomic and Molecular Sciences, Academia Sinica, Taipei 10617, Taiwan}

\author{H. H. Jen}%
\email{sappyjen@gmail.com}
\affiliation{Institute of Atomic and Molecular Sciences, Academia Sinica, Taipei 10617, Taiwan}
\affiliation{Molecular Science and Technology Program, Taiwan International Graduate Program, Academia Sinica, Taiwan}
\affiliation{Physics Division, National Center for Theoretical Sciences, Taipei 10617, Taiwan}

\date{\today}
\renewcommand{\r}{\mathbf{r}}
\newcommand{\f}{\mathbf{f}}
\renewcommand{\k}{\mathbf{k}}
\def\p{\mathbf{p}}
\def\q{\mathbf{q}}
\def\bea{\begin{eqnarray}}
\def\eea{\end{eqnarray}}
\def\ba{\begin{array}}
\def\ea{\end{array}}
\def\bdm{\begin{displaymath}}
\def\edm{\end{displaymath}}
\def\red{\color{red}}
\pacs{}
\begin{abstract}
Atomic array coupled to a one-dimensional nanophotonic waveguide allows photon-mediated dipole-dipole interactions and nonreciprocal decay channels, which hosts many intriguing quantum phenomena owing to its distinctive and emergent quantum correlations. In this atom-waveguide quantum system, we theoretically investigate the atomic excitation dynamics and its transport property, specifically at an interface of dissimilar atomic arrays with different interparticle distances. We find that the atomic excitation dynamics hugely depends on the interparticle distances of dissimilar arrays and the directionality of nonreciprocal couplings. By tuning these parameters, a dominant excitation reflection can be achieved at the interface of the arrays in the single excitation case. We further study two effects on the transport property-of external drive and of single excitation delocalization over multiple atoms, where we manifest a rich interplay between multi-site excitation and the relative phase in determining the transport properties. Finally, we present an intriguing trapping effect of atomic excitation by designing multiple zones of dissimilar arrays. Similar to the single excitations, multiple excitations are reflected from the array interfaces and trapped as well, although complete trapping of many excitations together is relatively challenging at long time due to a faster combined decay rate. Our results can provide insights to nonequilibrium quantum dynamics in dissimilar arrays and shed light on confining and controlling quantum registers useful for quantum information processing. 
\end{abstract}
\maketitle
\section{Introduction}

The fascinating realm of quantum nanophotonics \cite{Chang2018, Samutpraphoot2020, Sheremet2023} has witnessed significant advancements in recent years, owing to improving fabrication techniques and better control of light-matter interactions \cite{Vetsch2010, Thompson2013, Goban2015, Corzo2019, Kim2019, Dordevic2021}. This strongly-coupled atom-waveguide platform has shown many intriguing phenomena, including collective radiative dynamics \cite{Henriet2019, Zhang2019, Ke2019, Albrecht2019, Needham2019, Jen2020_subradiance, Mahmoodian2020, Masson2020, Kim2021, Jen2021_bound, Pennetta2022, Pennetta2022_2}, photon-mediated \cite{Zhong2020} or disorder-induced localization \cite{Mirza2017, Jen2020_disorder, Jen2021_localization, Fayard2021, Wu2023}, distinctive quantum correlations \cite{Tudela2013, Mahmoodian2018, Jeannic2021, Jen2022_correlation}, superior cooling behavior \cite{Wang2022, Chen2023}, and graph states generation \cite{Hiew2023, Chien2023}. At the heart of this burgeoning field of waveguide quantum electrodynamics is the long-range dipole-dipole interaction among the nanoscopic lattices of atoms mediated by the guided modes of the nanophotonic waveguide \cite{Douglas2015, Solano2017}. The effective long-range dipole-dipole interaction along with adaptable directional couplings between constituent atoms opens up a distinguishing field of chiral quantum optics \cite{Mitsch2014, Ramos2014, Pichler2015, Lodahl2017, Jen2020_phase} where the time-reversal symmetry is broken \cite{Bliokh2014, Bliokh2015}.

Chiral couplings refer to asymmetric decay channels among the atoms, which can transport the atomic excitations unidirectionally \cite{Gardiner1993, Carmichael1993, Stannigel2012, Jen2019_selective} and lead to nontrivial spin-exchange interactions or excitation interferences. This spin-momentum locking mechanism presents the essential element in chirally-coupled systems, which has been demonstrated experimentally in diverse platforms, for example the superconducting qubits \cite{Roushan2017, Wang2019}, quantum dots \cite{Luxmoore2013, Arcari2014, Yalla2014, Sollner2015}, trapped atoms \cite{Mitsch2014, Solano2017, Corzo2019, Morrissey2009, Tiecke2014, Sayrin2015}, and color centers \cite{Sipahigil2016, Bhaskar2017}. The directionality or the chirality of the couplings can be tailored by state-controlled spontaneous emissions \cite{Mitsch2014}, which gives rise to an extra degree of freedom in manifesting novel system dynamics, in addition to other variable factors like interparticle spacings, periodic or random distributions \cite{Song2021}, single or multiple excitations \cite{Jen2021_bound, Jen2022_correlation}.    

\begin{figure}[b]
\centering
\includegraphics[width=0.48\textwidth]{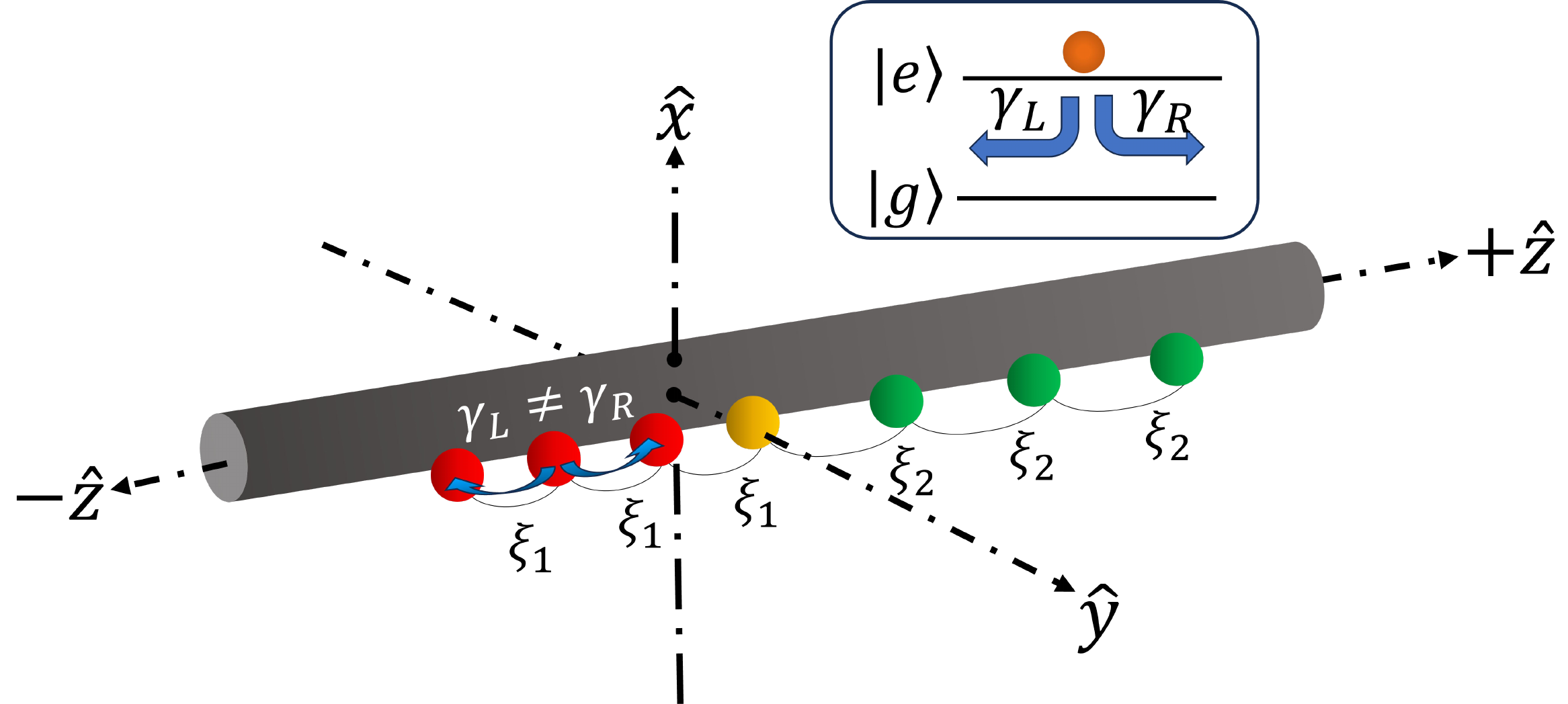}
\caption{Schematics of dissimilar chirally-coupled atomic arrays. The atomic chain is composed of two dissimilar atomic arrays with different interparticle spacings $\xi_1$ (red) and $\xi_2$ (green), respectively. An atom located at the interface of dissimilar arrays is denoted as the probe atom (yellow) with two disparate nearest-neighbor spacings. The inset plot shows a two-level quantum emitter ($\ket{g}$ and $\ket{e}$ indicating the ground and excited states, respectively) with nonreciprocal decay channels $\gamma_L \neq \gamma_R$ along the $\hat{z}$-axis.}\label{fig1}
\end{figure}

Among all these efforts to unravel exotic collective dynamics or to explore parameter regimes for novel applications, for instance in routing photons \cite{Sollner2015}, a recent endeavor tries to reveal the mechanism of quantum avalanche \cite{Leonard2023} as a signature of many-body delocalization \cite{Roeck2017, Luitz2017, Thiery2018, Morningstar2022, Sels2022} in an interface between the clean and disordered zones in an atomic array. This showcases the essence of an interface that bridges different physical domains, where distinct features of particle transport can emerge owing to the interferences between different zones and provide insights to the many-body localized phase \cite{Abanin2019} that is still under debate \cite{Lev2014, Panda2020, Abanin2021, Sierant2022}. Here we investigate the atomic excitation dynamics and excitation transport in a setup interfaced with dissimilar atomic arrays as shown in Fig. \ref{fig1}, which is less explored in open quantum systems. We find that the atomic excitation dynamics and excitation reflection or transmission hugely depend on the interparticle distances of dissimilar arrays and the directionality of nonreciprocal couplings. We further study the effect of multi-atom interference of single excitation on the transport property and present an intriguing excitation trapping effect in multiple zones of dissimilar arrays. Our results can provide insights to the role of an interface of dissimilar arrays in nonequilibrium quantum dynamics and potential applications of controlling and processing quantum information in a confined region. We note that the use of atomic arrays as optical
mirrors has been demonstrated \cite{Corzo2016}, and by exploiting this effect, optical cavities can be constructed using atomic chains \cite{Chang2012}. By contrast, here we focus on the atomic excitation dynamics and its transport property, where the excitation reflection effect we demonstrate is a physically distinct effect from photon reflections.

The remainder of the paper is organized as follows. In Sec. II, we introduce the theoretical model of dissimilar arrays characterized by two unequal interparticle distances with nonreciprocal couplings. In Sec. III, we study the single-site excitation dynamics and its transport property or penetration through the interface. We further reveal the multi-atom effect of single excitation in the excitation dynamics in Sec. IV and present the excitation trapping effect in multiple zones of dissimilar arrays in Sec. V. The effect of multiple atomic excitations in the trapping behaviors is also studied. Finally, we discuss and conclude in Sec. VII.   

\section{Theoretical model of dissimilar chirally-coupled atomic chain}

As shown in Fig. \ref{fig1}, we consider a system of dissimilar atomic arrays composed of the left and the right zones with equal sizes and interparticle spacings $d_{1,2}$, respectively, where we quantify them in terms of dimensionless $\xi_{1,2}\equiv k_sd_{1,2}$ with an atomic transition wavevector $k_s$. At the interface of two zones, we denote the atom as the probe atom which involves two disparate nearest-neighbor spacings $\xi_1\neq\xi_2$. We note that the spacings considered here are away from vanishing or short distances where the near-field effects may play an important role \cite{Kuraptsev2020}. Therefore, whenever the interparticle distances are set close to zero, they should be with an offset of a multiple of $2\pi$. Our theoretical model here follows closely to the conventional waveguide QED system \cite{Sheremet2023} where the influence of short-range interaction can be negligible. The density matrix $\rho$ of $N$ atoms with nonreciprocal decay channels in an interaction picture evolves as ($\hbar=1$) \cite{Pichler2015},
\bea
\frac{d\rho}{dt}=-i[H_L +H_R,\rho] + \mathcal{L}_L[\rho] + \mathcal{L}_R[\rho],\label{rho}
\eea
where the coherent and dissipative system dynamics are determined by
\bea
H_{L(R)}=-i\frac{\gamma_{L(R)}}{2}\sum_{\mu<(>)\nu}^N\sum_{\nu=1}^N (e^{ik_s| r_\mu - r_\nu |}\sigma_\mu^\dagger\sigma_\nu - \rm{H.c.})\label{H}
\eea
and
\bea
\mathcal{L}_{L(R)}[\rho]=&&-\frac{\gamma_{L(R)}}{2}\sum_{\mu,\nu}^N e^{\mp ik_s(r_\mu - r_\nu)}(\sigma_\mu^\dagger\sigma_\nu\rho + \rho \sigma_\mu^\dagger\sigma_\nu \nonumber\\
&&- 2\sigma_\nu\rho\sigma_\mu^\dagger)\label{L}.
\eea
The dipole operators are $\sigma_\mu^\dagger\equiv\vert e \rangle_\mu\langle g\vert$ with $\sigma_\mu = (\sigma_\mu^\dagger)^\dagger$, and the nonreciprocal coupling strengths are denoted by $\gamma_{L}\neq \gamma_{R}$. The above effective density matrix equation is derived under the Born-Markov approximation \cite{Lehmberg1970} and 1D reservoirs \cite{Tudela2013}, where an infinite-range photon-mediated dipole-dipole interaction \cite{Mitsch2014} can be supported by the guided modes on the waveguide \cite{Sheremet2023}. 

In Eq. (\ref{rho}), we can order the atomic positions as $r_1 < r_2 < \cdots <r_{N-1} < r_N$, where we define the atomic spacing as $d_{1,2} = r_{\mu + 1} - r_\mu$, where the spacing is $d_1$ or $d_2$ if the $\mu$th atom is located in the left or the right zone. 

We also define the probe atom, as the atom at the interface between two zones with equal sizes, and denote its site index as $n_p$. The tendency of effective excitation transfer can be quantified as the directionality factor $D\equiv (\gamma_R - \gamma_L)/\gamma$ \cite{Mitsch2014} with $\gamma = \gamma_R + \gamma_L \equiv 2\vert dq(\omega)/d\omega\vert_{\omega = \omega_{eg}}g^2_{k_s}L$ \cite{Tudela2013}. $\vert dq(\omega)/d\omega\vert$ denotes the inverse of group velocity with a resonant wavevector $q(\omega)$, $g_{k_s}$ is the atom-waveguide coupling strength, and $L$ denotes the quantization length. $D\in [-1,1]$ specifies the trend of photon exchange between quantum emitters, where $D = \pm 1$ present the unidirectional coupling scheme \cite{Carmichael1993, Gardiner1993, Stannigel2012}, while $D = 0$ represents a reciprocal coupling regime with equal decay rates.

We consider to initialize the system from a single atomic excitation either in the left or the right zone, which preserves the excitation number, and a single excitation subspace $\vert \psi_\mu \rangle = \vert e \rangle_\mu \vert g \rangle^{\otimes(N-1)}$ should be sufficient to describe the system dynamics. Within this subspace, the general dynamical equations of a density matrix can be reduced to a non-Hermitian Schr\"{o}dinger equation with an effective Hamiltonian $H_{\rm eff}$,   
\begin{eqnarray}
i\frac{\partial}{\partial t}\vert \Psi(t)\rangle = H_{\rm eff}\vert \Psi(t)\rangle, \label{Psi}
\end{eqnarray}
where the state of the system is $\vert \Psi(t)\rangle =\sum_{\mu = 1}^N a_\mu (t)\vert \psi_\mu\rangle$ with the probability amplitudes $a_\mu (t)$. Then, we obtain the coupled equations of the system as
\begin{eqnarray}
\dot{a}_\mu(t) = &&-\gamma_R\sum_{\nu<\mu}^N e^{-ik_s(r_\mu-r_\nu)}a_\nu(t)-\frac{\gamma}{2}a_\mu(t)\nonumber\\
&&-\gamma_L\sum_{\nu>\mu}^N e^{-ik_s(r_\nu-r_\mu)}a_\nu(t).
\end{eqnarray}
In the below, we analyze the system dynamics and the excitation transport property by solving the above equations, and further identify several interesting parameter regimes that can host dominant excitation reflection and excitation trapping in a design of multiple zones. 

\section{Atomic excitation dynamics and transport}

\begin{figure}[tb]
\includegraphics[width=0.48\textwidth]{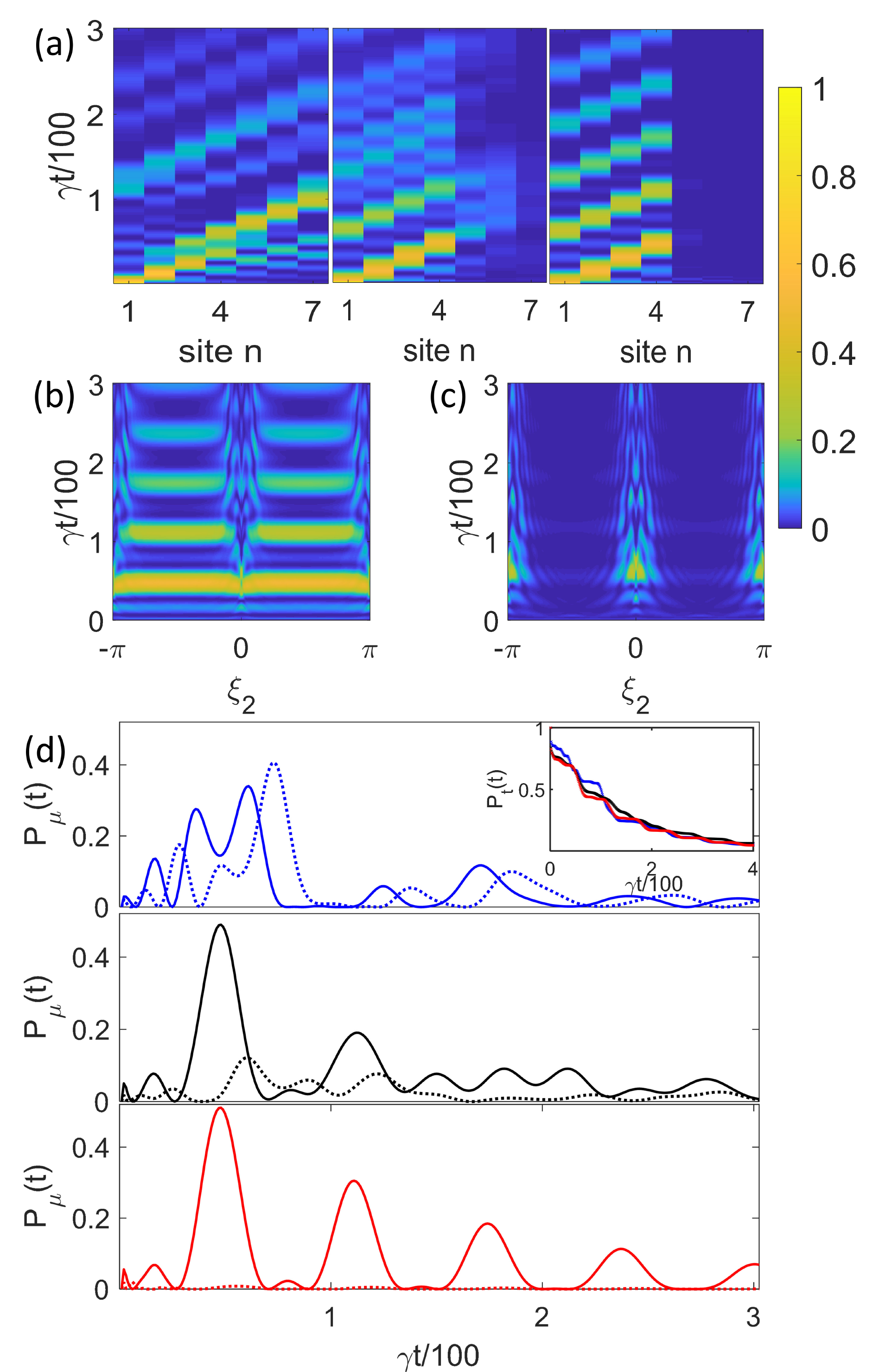}
\caption{\label{fig2} A dramatic transition to near total excitation reflection at the interface with changing $\xi_2$. In all plots, we calculate the time evolution of population $P_n(t)$ at $D=0.2$ and $\xi_1=\pi$ for $N = 7$. Under a fixed $\xi_1=\pi$, the excitation dynamics are shown for (a) $\xi_2=\pi$ (left), $\pi/8$ (middle), and $\pi/2$ (right), respectively. The population of (b) the probe atom $P_4(t)$ and (c) the atom $P_5(t)$ at the site $n=5$ versus various $\xi_2$ at $\xi_1=\pi$. (d) Decays of the probe atom (solid-blue, solid-black, and solid-red lines) and the nearest-neighbor atom $P_5(t)$ (dashed-blue, dashed-black, and dashed-red lines) for $\xi_2=\pi$, $\pi/8$, and $\pi/2$, respectively. The inset plot presents the corresponding total population decay for $\xi_2=\pi$ (solid-blue line), $\pi/8$ (solid-black line), and $\pi/2$ (solid-red line), respectively.}
\end{figure}

The excitation transport and decay behavior of atoms coupled with a one-dimensional waveguide represents a fascinating and pivotal area of study in waveguide QED \cite{Masson2020, Sheremet2023}. Here we focus on the setup with dissimilar atomic arrays and investigate the decay behavior of these quantum emitters under nonreciprocal couplings when a single atomic excitation is initialized. In Fig. \ref{fig2}, we take $N=7$ as an example, where the probe atom sits at the site of $n_p=4$. We set the initial excitation at the first site $n=1$, and we trace the individual atomic populations until the total population reaches around $10\%$.

In Fig. \ref{fig2}(a), we plot the homogenous case where $\xi_1=\xi_2=\pi$ as a comparison in the left subplot, which is the regime of subradiant decay that allows long-time decay behavior \cite{Jen2020_subradiance}. The excitation transfers preferentially to the right owing to the spin-exchange couplings determined by $\gamma_{L(R)}$ and the chosen $D>0$. The primary populations of $P_n(t)=|a_n(t)|^2$ undergo a ballistic diffusion as they move towards the lattice boundary. A repopulation occurs on the other side of the lattice boundary, and the excitations continue to propagate with population interferences surrounding the main population. The former is due to the feature of infinite-range dipole-dipole interactions that can transfer and exchange the populations at both boundaries of the lattices, while the latter can be attributed to the finite $D<1$ and counter-propagating decay channels. As $\xi_2$ approaches $\pi/2$, the excitation penetration to the right zone diminishes and indicates an almost reflected population at the interface. Essentially, the repopulation only accumulates in the left zone. 

We further investigate the time dynamics of $P_4(t)$ and $P_5(t)$ for various $\xi_2$ in Figs. \ref{fig2}(b) and \ref{fig2}(c), respectively, which can reveal how the excitation population transports through them in time. As expected, a symmetric profile of excitation population within $-\pi \leq \xi_2 \leq \pi$ is observed around $\xi_2 = 0$, owing to the mirror symmetry in $\xi_2$ preserved in the open waveguide setup. In particular, the probe atom experiences a periodic pattern of repopulation when $\xi_2$ approaches and narrows around $\pi/2$ as time evolves. This periodic but decaying oscillation for the probe atom is more evident in Fig. \ref{fig2}(d), where a suppression of $P_5(t)$ can be manifested as well, making a halt of the population transfer at the interface. By contrast, $P_5(t)$ increases significantly only at a short time when $\xi_2$ is around $\pi$ or $\pi/8$, and in this regime of interparticle spacing, both the probe atom and the atom next to it in the right zone decay and exchange populations sequentially in time. We note that the total excited-state populations are almost the same for different cases as time evolves, which is shown in the inset plot of Fig. \ref{fig2}(d). Therefore, the role of the interface indeed redirects the excitation populations and make them more concentrated in the left zone. 

The periodic and reflection-dominant behaviors of the excitation dynamics in this atom-waveguide interface are a result of the long-range dipole-dipole interactions and the specific values of $\xi_1$ and $\xi_2$ in the design of dissimilar arrays. The preceding findings offer valuable insights into the intricate relationship between the atomic configurations and the excitation processes in waveguide-coupled systems. Nevertheless, the observation that an almost complete excitation reflection takes place when $\xi_1 = \pi$ and $\xi_2 = \pi/2$ is not applicable in the case of large $D$. For instance, in the unidirectional case ($D=1$), the population predominantly propagates only in the right direction regardless of any atomic spacings, and the long-term survival of the population is not feasible \cite{Jen2020_collective}. 

\begin{figure}[tb]
\includegraphics[width=0.48\textwidth]{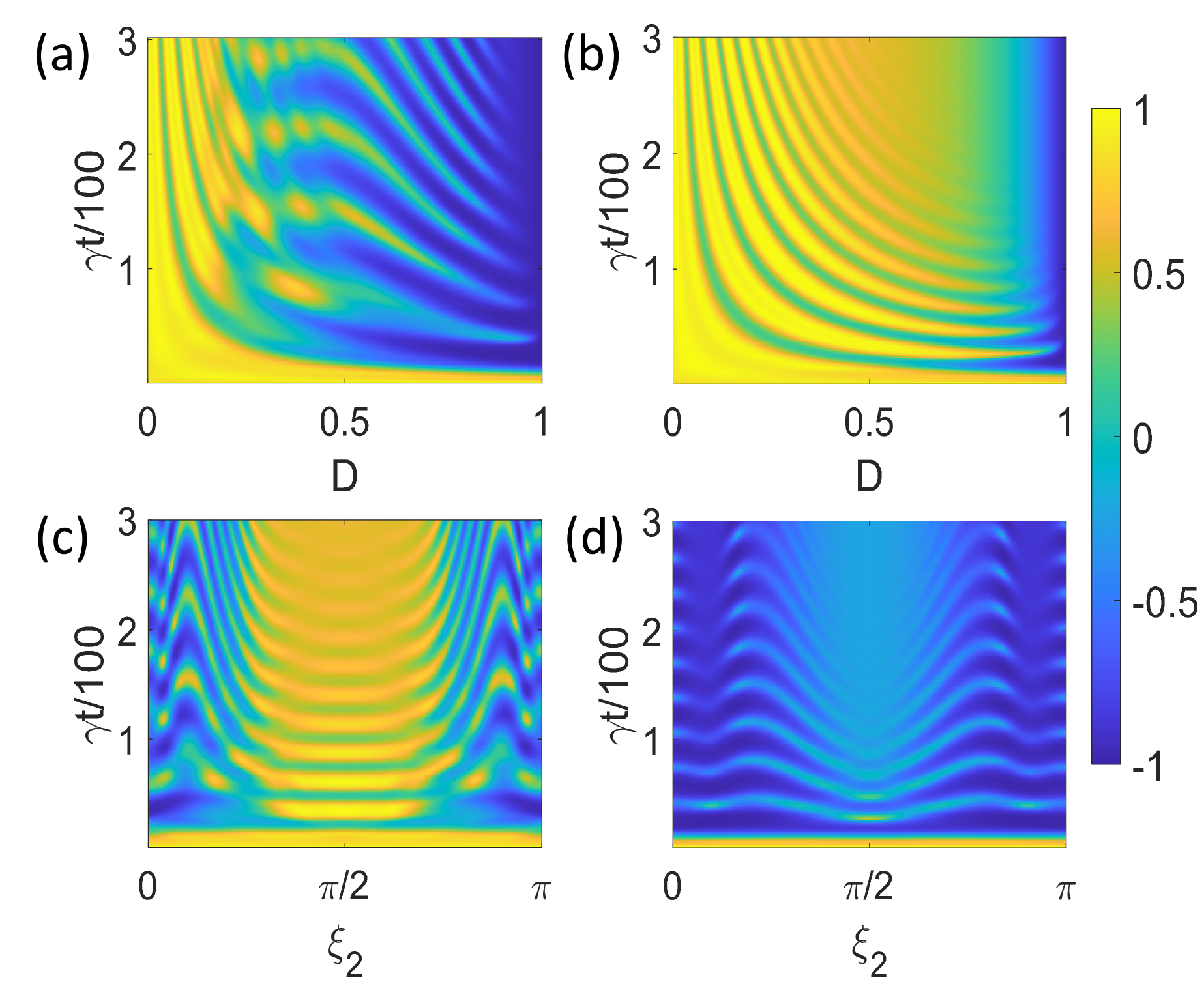}
\caption{The transport parameter $T_p(t)$ [defined in Eq. (\ref{Tp})] is plotted to show the competition between $D$ and $\xi_2$ on the atomic excitation dynamics (i.e., reflection or transmission) at the interface. For finite $D$, $T_p(t)$ remaining close to $1$ over time generally signifies the excitation reflection from the interface, while a change to $-1$ signifies transmission. $T_p (t)$ from a single-site atomic excitation for $N = 7$ is plotted for various $D$ at (a) $\xi_2=\pi/8$ and (b) $\pi/2$, and for various $\xi_2$ at (c) $D=0.5$ and (d) $0.9$, under a fixed $\xi_1=\pi$.}\label{fig3}
\end{figure}

To reveal the effect of directionality factor $D$ on the atomic excitation distributions, we define the transport parameter $T_p$ by the difference of excited-state populations between the left and right sections of the atomic array \cite{Jen2019_selective}, which for an even or odd $N$ can be expressed as, 
\begin{equation}
T_p = \frac{\Sigma^{N/2, (N-1)/2}_{\mu=1}P_\mu(t)-\Sigma^{N}_{\mu=N/2+1, (N+3)/2}P_\mu(t)}{\Sigma^N_{\mu=1}P_\mu(t)},\label{Tp}
\end{equation}
where we exclude the probe atom with an odd $N$ for a balance of number of atoms in obtaining the transport parameter. Again we take $N = 7$ as an example in Fig. \ref{fig3} and show the transport parameter for various $D$ and $\xi_2$ at the same $\xi_1$ in Fig. \ref{fig2}. In Figs. \ref{fig3}(a) and \ref{fig3}(b), the effect of a larger $D$ drives $T_p$ toward negative values, showing the influence of excitation penetration into the right zone of the dissimilar arrays. Most of the excited-state populations stay in the left zone when the dominant reflection emerges for a smaller $D$. The optimal values of $\xi_2$ for significant excitation reflection can be identified as well in Fig. \ref{fig3}(c) for a higher $D$ than in Fig. \ref{fig2}, where the stripe patterns in $T_p$ coincide with a halt of atomic excitation at the probe atom when $T_p$ vanishes and with the periodic oscillations in Fig. \ref{fig2}(a). Figure \ref{fig3}(d) under a $D$ close to the unidirectional coupling regime presents the case with a dominant transmission instead, where most of $T_p$ is less than zero.  

\begin{figure}[tb]
\includegraphics[width=0.48\textwidth]{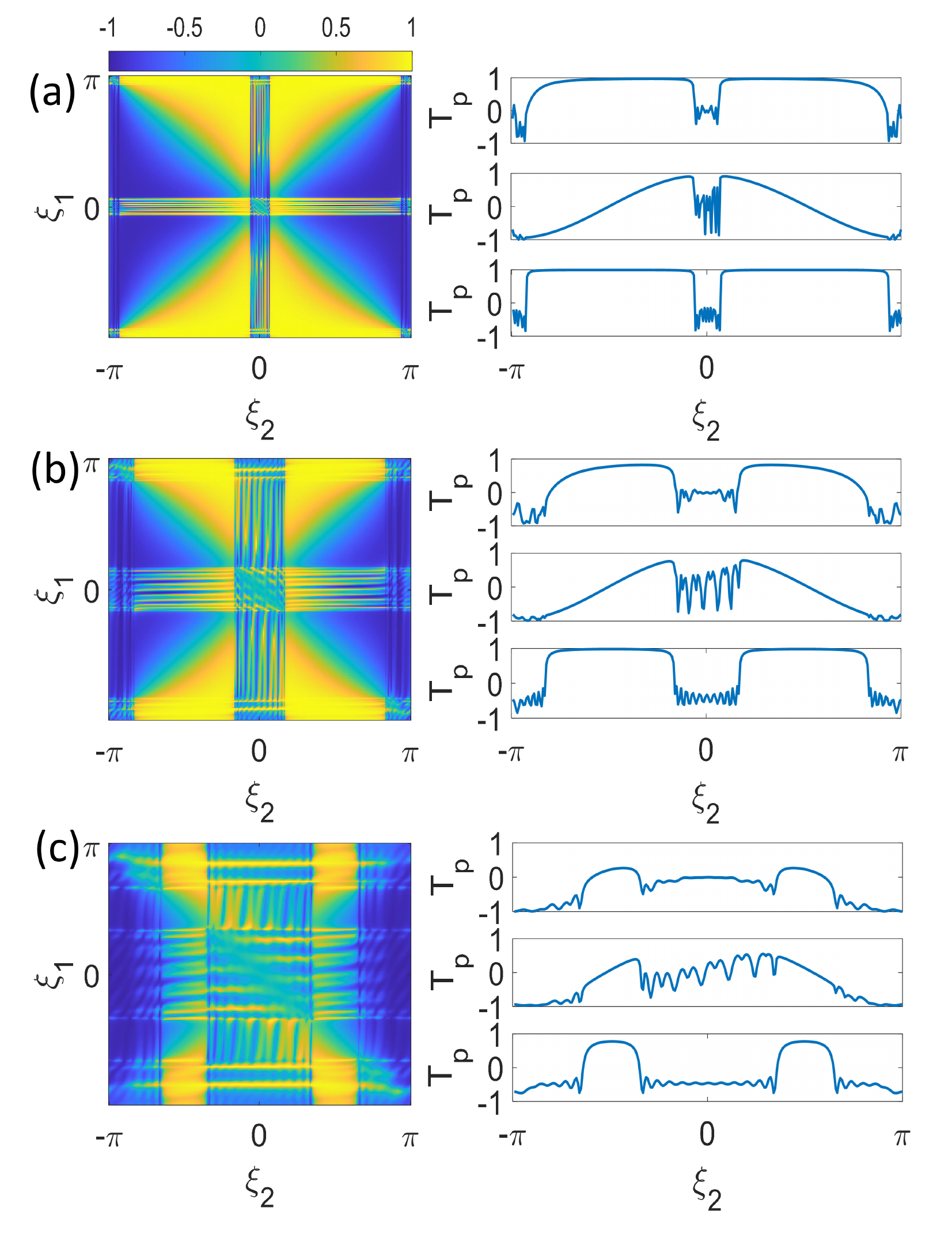}
\caption{Symmetric transport parameters of a weakly-driven dissimilar atomic arrays with $\xi_1$ and $\xi_2$. In the steady-state solutions for $N = 24$, we obtain $T_p$ at (a) $D=0.2$, (b) $0.5$, and (c) $0.9$ for various $\xi_1$ and $\xi_2$ in the left panels. Several cross-section profiles in the corresponding right panels are demonstrated for $\xi_1 = 0$ (upper subplot), $\pi/2$ (middle), and $\pi$ (bottom), respectively.}\label{fig4}
\end{figure}

As a comparison, we study the transport parameter for dissimilar atomic arrays under a weakly-driven and uniform laser excitation. From the steady-state solutions, we calculate the distributions of the atomic excited-state populations as shown in Fig. \ref{fig4}. The $T_p$ shows symmetric patterns around $\xi_{2(1)}=0$ when $\xi_{1(2)}=0$ or $\pi$ for all $D$ considered in Fig. \ref{fig4}. This symmetry should be applied to all other fixed $\xi_1$ or $\xi_2$ when the system reaches the thermodynamic limit. Within $-\pi<\xi_{1(2)}<\pi$, we find homogeneous distributions between two zones when $\xi_1\approx\pm \xi_2$, corresponding to $T_p \approx 0$. From the cross-section profiles, we locate the highest $T_p\approx 1$ which occurs when $\xi_1 = \pm\pi$ and $\xi_2 = \pm\pi/2$. This condition again coincides with the one applied in the excitation dynamics when significant excitation reflection emerges as discussed in Figs. \ref{fig2} and \ref{fig3}. As a final remark, there are wide ranges of system parameters that allow significant reflection behaviors, but these regions shrink as $D$ increases as shown in Fig. \ref{fig4}. This presents the competition between the facilitation of halting atomic excitations at the interface from interferences of long-range spin-exchange interactions and the driving force to penetrate the interface augmented by a large $D$.  

\section{Multi-site excitation dynamics}

Next, we study the excitation transport from a single atomic excitation initialized at two sites, as an extension to the results in the previous section. For two sites excitation, we use the initial state $a(0) = (\sigma_1^\dagger + e^{i\theta}\sigma_2^\dagger)\vert g \rangle^{\otimes N}/\sqrt{2}$ with a dependence of a relative phase $\theta$, and for multiple sites we consider a collective state $a(0) = (\sum_{\mu=1}^M e^{i\theta\delta_{\mu,M}}\sigma_\mu^\dagger \vert g \rangle^{\otimes N})/\sqrt{M}$ with a phase dependence on the Mth site as a comparison. When $\theta=0$, these multi-site excitation states become one of the Dicke's symmetric states \cite{Dicke1954}. In Sec. III, it is evident that the excitation transport demonstrates an optimal reflection at the interface of two zones when the interparticle spacings are $(\xi_1,\xi_2)=(\pi,\pi/2)$. This observation holds true even for a two-site or multi-site initialized states with arbitrary $\theta$. Meanwhile, the optimal excitation reflection from the interface can also be observed for many other combinations of $(\xi_1,\xi_2)$ with $\theta$ playing a significant role in determining the dominant excitation reflection. 

\begin{figure}[tb]
\includegraphics[width=0.48\textwidth]{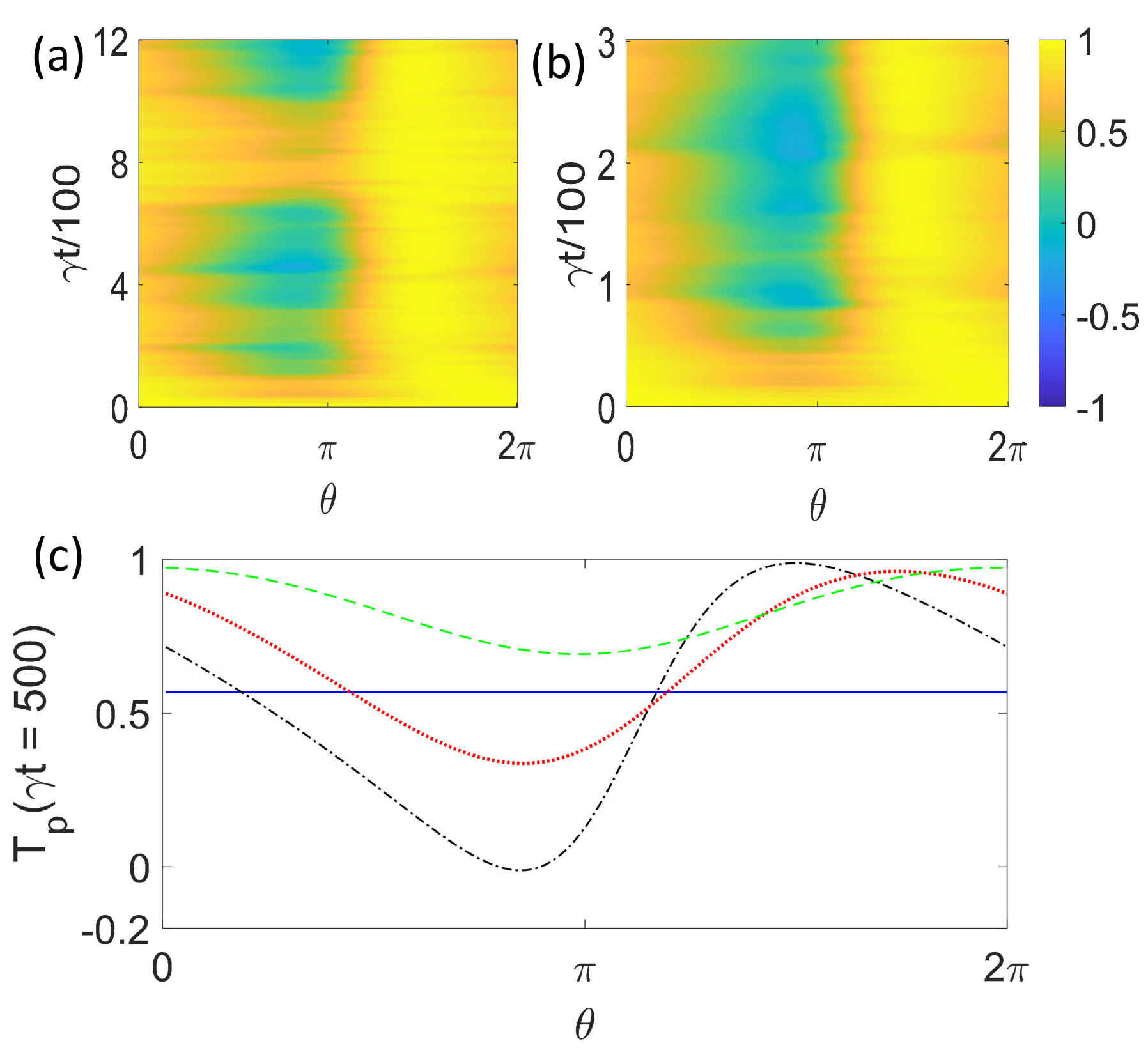}
\caption{Time-evolving transport parameter $T_p (t)$ with a multi-site atomic excitation for $N=24$. The dynamics of a single excitation delocalized over multiple atoms is different from a single-atom excitation described before. By tuning the phase between the atoms the transport parameter $T_p (t)$ can be altered. For two-site excitation at the site $n=1$ and $2$ with a dependence of $\theta$, we plot $T_p$ for (a) $D=0.2$ at $(\xi_1,\xi_2) = (\pi,\pi/8)$ and (b) $D = 0.4$ at $(\xi_1,\xi_2) = (\pi,\pi/4)$, respectively. (c) For multi-site excitation, we compare $T_p$ for $M=1$ (solid-blue line), $2$ (dash-dotted line in black), $3$ (dotted-red line), and $4$ (dash-green line), under the same system parameters in (a).}\label{fig5}
\end{figure}

In Fig. \ref{fig5}, we present two cases of $D$, which illustrate the transport parameter $T_p$ for two-site excitation at $(\xi_1,\xi_2)=(\pi,\pi/8)$ and $(\xi_1,\xi_2)=(\pi,\pi/4)$, respectively. With $\theta$ extending from $0$ to $2\pi$, we present $T_p$ in time until a total excited-state population decays to the $10\%$ of the initialized state. It is evident that when $\theta\lesssim \pi$, most of the population tends to propagate towards the right zone comparing to the case of $\theta\approx 0,~2\pi$. This tendency is particularly pronounced for certain values of $\xi_2$, specifically $\xi_2 \approx \pi/8$ or $\pi/4$ as shown in Figs. \ref{fig5}(a) and \ref{fig5}(b). By contrast, when $\theta > \pi$, a significant transition towards a dominant reflection occurs, where $T_p$ approaches unity. For a multi-site excitation setup considered in Fig. \ref{fig5}(c), we find the trend of multi-atom effect which enhances $T_p$ as $M$ increases, whereas for a finite $M<4$, a contrasted $T_p$ can be seen as $\theta$ changes. This relates the information encoded in $\theta$ of the initialized states to its transport behaviors and potentially provides a mechanism for routing quantum information in the atom-waveguide interface. Figure \ref{fig5} shows an example of rich interplay between multi-site excitation and the relative phase in determining the transport properties. We expect new results would emerge as full-fledged involvement of all degrees of freedom in multi-site excitation can be explored.   

\section{Excitation trapping in multiple zones}

\begin{figure}[t]
\includegraphics[width=0.48\textwidth]{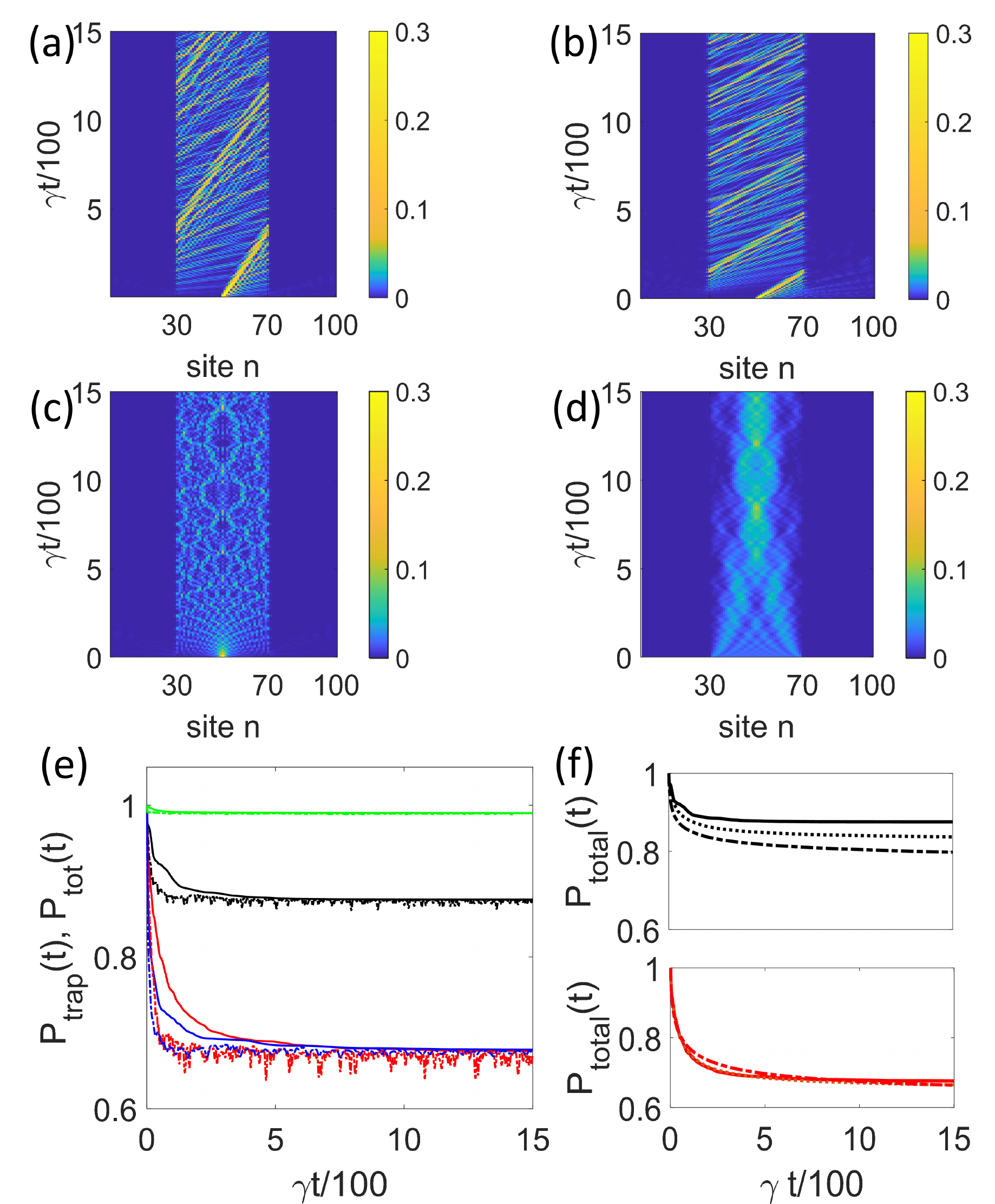}
\caption{Excitation trapping effect in dissimilar arrays. The atomic excitation can be confined by introducing two interfaces with a design of three zones with $\xi_{1,2,3} = \pi/2$, $\pi$, $\pi/2$ (left, middle, and right zones) for $N = 100$, $N_{1,2,3} = 30,~40,~30$, respectively, where two probe atoms at the interface of sites $n = 30$ and $70$. We demonstrate the excited-state populations $P_\mu(t)$ for (a) $D = 0.2$ (b) $D = 0.5$ under a single-site excitation in the center of the middle zone. In (c), we study the case at $D=0$ with $\xi_2 = \pi/8$ instead under a single-site excitation. (d) For the multi-site excitation, we initialize the system throughout the sites from $n=30$ to $70$, forming a symmetric Dicke's state, with $\xi_{1,2,3} = \pi/2, 3\pi/4, \pi/2$. (e) We compare the total excited-state populations (solid line) with the populations trapped in the middle zone (dashed line) for the cases in (a), (b), (c), and (d), denoted by the colors in black, red, blue, and green, respectively. The total population $P_{\rm tot}(t)$ for each case saturates and remains constant mainly contributed by the population in the trapping zone $P_{\rm trap}(t)$. (f) The total population in excitation trapping phenomena decreases while position fluctuation is added by $1\%$ (dotted line) and $5\%$ (dash-dotted line) comparing with no deviation (solid line), shown for the cases (a) and (b) in the upper and lower panel, respectively}\label{fig6}
\end{figure}

Here we investigate the excitation trapping effect when an extra zone is introduced into the dissimilar arrays we consider in the previous sections. Based on the system parameter regimes that illustrate a dominant excitation reflection, we design a cavity-like setup that can confine the initialized atomic excitation to achieve precise control and manipulation of excitation transport. This leads to a setup of three zones in a dissimilar array with number of atoms $N_1$, $N_2$, and $N_3=N_1$, and interparticle spacings $\xi_1$, $\xi_2$, and $\xi_3=\xi_1$. As shown in Fig. \ref{fig6}, the middle zone is delimited by two probe atoms, where we show the time dynamics of trapped atomic excitations from a single-site in Figs. \ref{fig6}(a), \ref{fig6}(b), and \ref{fig6}(c), or multiple sites in Fig. \ref{fig6}(d).

For $D>0$ in Figs. \ref{fig6}(a) and \ref{fig6}(b), the excitation propagates preferentially to the right of the middle zone with proportional ballistic speed to the values of directionality factors $D$. This trapping effect is expected when we use the parameters that manifest dominant excitation reflection. More interestingly, for $D=0$ in the reciprocal coupling regime, we observe two contrasting excitation dynamics from a single-site or multi-site initializations in Figs. \ref{fig6}(c) and \ref{fig6}(d), respectively. Both cases involve interference patterns in populations owing to the spin-exchange couplings, while the initialized symmetric Dicke state in Fig. \ref{fig6}(d) presents an agglomerated feature of population as time evolves. The trapping effect can be further evidenced by comparing the total populations $P_{\rm tot}(t)$ with the populations in the trapping zone $P_{\rm trap}(t)$ as shown in Fig. \ref{fig6}(e), where $P_{\rm tot}(t)$ is mainly contributed by $P_{\rm trap}(t)$. Furthermore, $P_{\rm trap}(t)$ from a multi-site excitation presents a relatively high total population, which indicates the multiatom effect that preserves the atomic excitation and is consistent with the enhancement of excitation reflection in Fig. \ref{fig5}(c). We note that there are slight ups and downs in $P_{\rm trap}(t)$ as it approaches $P_{\rm tot}(t)$. These fluctuations indicate the exchanging process of excitations at the interfaces, where the excited-state population leaking out from the interface can reenter into the middle zone. This however is not significant when the multi-site configuration is considered, which can be attributed to the initialized quantum correlations among the atoms, leading to a sustaining and high excitation population in the trapping zone.

In examining the impact of positional deviation, our study extends to simulating the effect of excitation trapping under atomic position fluctuations. This disorder effect can be unavoidable in the presence of local inhomogeneous fields in optical lattices or thermal vibrations, leading to possible deviations from the ideal geometry under consideration. As shown in Fig. \ref{fig6}(f), we perform the simulations for scenarios with $D = 0.2$ and $D = 0.5$, which we compare with the examples shown in Figs. \ref{fig6}(a) and \ref{fig6}(b), respectively. The presence of deviation in the atomic position weakens the trapping phenomena especially in the low $D$ case and tends to reduce the total population in the trapping area. This makes sense since the trapping effect is more significant at a low $D$ and is thus more fragile to the position fluctuations. We note that as the fluctuations increase, the localization of the excitation population emerges \cite{Jen2020_disorder} and dominates over the influence of excitation interferences determined by interparticle distances. 

Finally, a question arises as how many atoms needed in the trapping zone to preserve sufficient atomic excitations? We qualitatively determine the trapping effect in the atomic excitation whenever $P_{\rm tot}(t)>0.1$. Under this condition, we define the trend parameter as the normalized change of populations at a long time, which reads $[P_{\rm tot}(\gamma t=1000)-P_{\rm tot}(\gamma t=4000)]/P_{\rm tot}(\gamma t=1000)$. This distinguishes the trend or a slope of one when the trapping effect disappears from a vanishing slope when the excitation population stays constant as time evolves. The references of time are adopted from the case of $D=0.5$ with $N_1 = N_3 = 3$ and $N_2 = 1$, where its total population at $\gamma t = 1000$ is more than $10\%$, while at $\gamma t = 4000$ it becomes less than $1\%$. This shows a qualitative measure for the sustaining excitation population or not at a long enough time. In this study, the referenced case for $N_1 = N_2=N_3 = 1$ leading to a total number of atoms $N=3$ presents a homogeneous atomic array with an interparticle spacing $\xi_2$ only. Two probe atoms at two interfaces emerge only when $N_1=N_3>1$, which enables the effect of dissimilar arrays with different spacings $\xi_1=\xi_3$. 

\begin{figure}[t]
\includegraphics[width=0.48\textwidth]{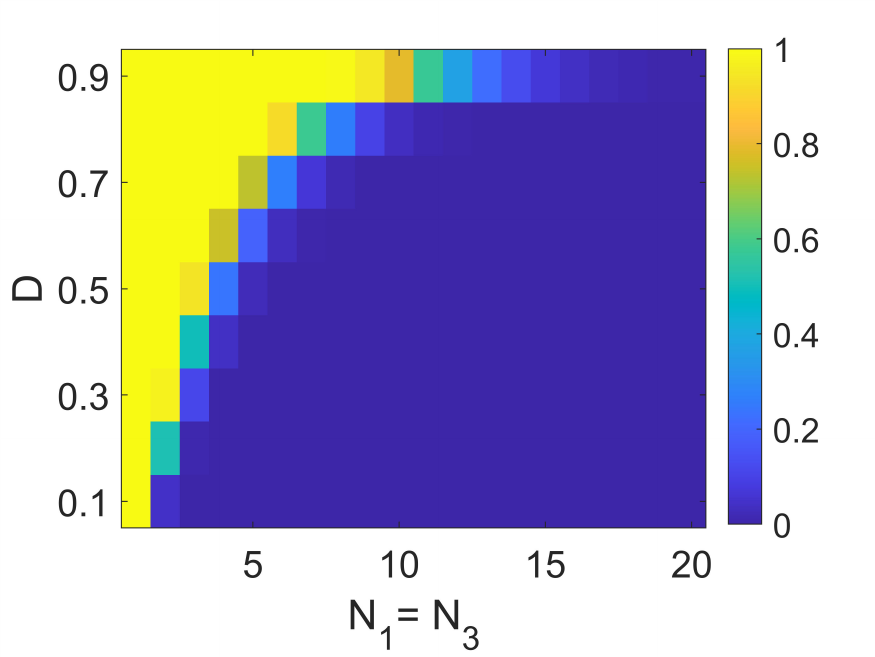}
\caption{\label{fig7}A trend of minimal number of atoms in the side zones ($N_1 = N_3$) required to preserve the atomic excitation in the middle zone of $N_2=1$ for various $D$. The other parameters are the same as in Figs. \ref{fig6}(a) and \ref{fig6}(b).}
\end{figure}

Figure \ref{fig7} shows the minimal number of atoms to maintain the excited-state population in the side zones for different $D$. As $D$ increases, more number of atoms are required to sustain sufficient excitation populations. For instance at $D = 0.5$, the minimal number of atoms required is six atoms to preserve a significant population at long time, while it takes three atoms in the case of $D = 0.2$. This figure of merit qualitatively shows the minimal resource of the number of atoms in the trapping zone to effectively retain the system population, which can be useful for storing quantum information or quantum information processing. We exclude the unidirectional ($D=1$) and reciprocal cases ($D=0$) since the former always leads to a penetration through the interface no matter what combinations of $\xi_{1,2,3}$ are, while the latter does not show trapping effect in the considered parameter regimes since it is always trapped owing to the decoherence-free space.

\section{Trapping effect of multiple atomic excitations}

\begin{figure}[b]
\includegraphics[width=0.48\textwidth]{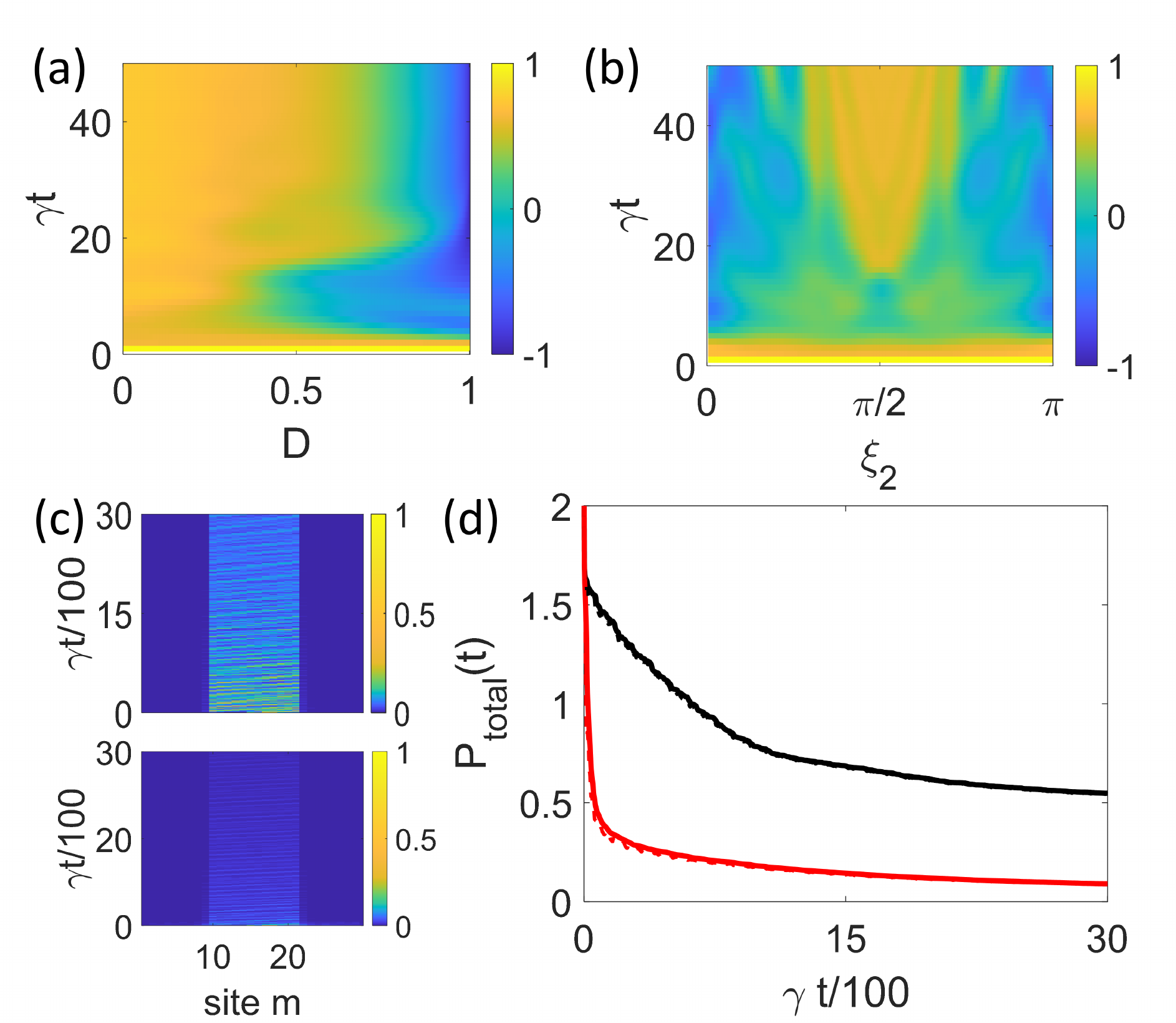}
\caption{Excitation trapping effect for double excitations. The time-evolution transport parameter $T_p(t)$ quantifies the reflection and transmission at the interface of the whole system, where we place two excitations initialized in the first two sites. All parameters used in (a) and (b) are the same as in Figs. \ref{fig3}(b) at $\xi_2=\pi/2$ and \ref{fig3}(c) at $D=0.5$, respectively, under a fixed $\xi_1=\pi$. (c) Excitation trapping effect for $N_1 = N_2 = N_3 = 10$, with the same parameters of $\xi_{1,2,3}$ and $D$ as in Figs. \ref{fig6}(a) and \ref{fig6}(b) in the upper and lower panels, respectively. (d) The total populations (solid line) and trapping populations (dashed line) for the cases in the upper and lower panels of (c) (black and red solid lines, respectively).}\label{fig8}
\end{figure}

Furthermore, we study the cases of multiple atomic excitations and investigate their reflection and trapping effects as a comparison and extension to the single excitation results in the previous sections. We utilize the construction of Hilbert space for multiply-excited states \cite{Jen2021_bound, Jen2022_correlation} and numerically simulate the transport parameter $T_p$ and time-evolving state populations. For $M$ atomic excitations out of $N\geq M$ atoms, we have a Hilbert space up to a number of a binomial coefficient $C^N_M$. This would grow exponentially with an order of $\mathcal{O}(N^M)$ when $N\gg M\gg 1$. On the other hand, the intrinsic decay rate with multiple excitations is $M$ times the single excitation decay. This indicates a faster deterioration for the trapping phenomena we observe in singly-excited systems. An essential element in this study focuses on the durability and effectiveness of the trapping phenomenon and one objective is to investigate if this phenomenon can maintain a finite number of confined excitations for long periods of time.

As shown in Figs. \ref{fig8}(a) and \ref{fig8}(b) for double atomic excitations initialized in the first and second sites, we calculate the time-evolving $T_p$ similar to Fig. \ref{fig3} and find that the reflection occurs at low $D$ and around $\xi_2=\pi/2$, respectively. One significant deviation from single excitation case is the limited time allowed for the total population to be sustained within $10\%$ compared to Fig. \ref{fig3}, owing to the enhanced intrinsic decays. In Fig. \ref{fig8}(c), we present the excitation trapping effect for $N_1 = N_2 = N_3 = 10$ with the same parameters as in Fig. \ref{fig6}. The corresponding atomic excited-state populations in the trapping zone as shown in Fig. \ref{fig8}(d) are relatively lower than the single excitation case. For higher $M$ excitations, the trapping effect becomes less significant and cannot last to long time, which is limited by the intrinsic decays. 

\section{Discussion and conclusion}

The coupling of dissimilar atomic arrays to a waveguide offers an unprecedented mechanism for controlling and manipulating excitation transport within quantum systems. In this quantum interface with controlled zone parameters, a dominant excitation reflection and a retention of excitations enable an efficient transfer of quantum information between separate zones. This manipulation ultimately leads to the emergence of a trapping effect which we present here. This not only offers a paradigm for studying nonequilibrium quantum dynamics, but also unique opportunities in controlling stationary qubits, essential in applications of quantum technology. Several potential platforms are available for experimentally observing the excitation dynamics and trapping effect, for example a photonic crystal setup \cite{Arcari2014, Goban2015} or an atom-fiber system \cite{Corzo2019}. In particular, optical tweezers can be applied to further enhance the control of the atoms \cite{Kim2019, Samutpraphoot2020, Dordevic2021}, reaching the strong coupling regime in the atom-waveguide interface \cite{Samutpraphoot2020}.

With the capability of trapping atomic excitations and utilizing the dissimilar arrays associated with excitation transport behaviors, a quantum storage and retrieval of qubit information can be facilitated by adiabatically modifying the dissimilar configurations to the homogeneous one. This enables the development of highly efficient and low-noise device as quantum memory. Our results promise to shed light on the complex interplay between atomic arrangements and collective spin-exchange interactions, offering insights into how we can engineer light-matter interactions. As the field of chirally-coupled atomic systems continues to evolve, future research may include refining the design of atomic arrangements to generate graph states for problem-specific applications \cite{Chien2023}, optimizing interatomic spacings, and exploring novel quantum phenomena that can be engineered through chirality of the couplings. 

\section*{ACKNOWLEDGMENTS}
We acknowledge support from the National Science and Technology Council (NSTC), Taiwan, under Grants No. 112-2112-M-001-079-MY3 and No. NSTC-112-2119-M-001-007, and from Academia Sinica under Grant AS-CDA-113-M04. We are also grateful for support from TG 1.2 of NCTS at NTU.

\end{document}